\documentclass[conference]{HPCSTrans}
\IEEEoverridecommandlockouts
\usepackage{amsmath,amssymb,enumerate}
\usepackage{cite,graphics,graphicx}
\usepackage{graphicx}
\usepackage{subfigure}
\usepackage{makeidx}
\usepackage{hyperref}
\usepackage{algorithm}
\usepackage{algorithmic}

\usepackage{multirow}
\usepackage{amsmath}
\usepackage{amssymb}
\usepackage{xcolor}
\usepackage{epsfig}
\usepackage{graphics}
\usepackage{listings}

\begin{document}
%
\title{Ultra-fast Multiple Genome Sequence Matching Using GPU}



%
\author{\IEEEauthorblockN{Gang Liao\IEEEauthorrefmark{1}\IEEEauthorrefmark{3},
Qi Sun\IEEEauthorrefmark{1},
Longfei Ma\IEEEauthorrefmark{1},
Sha Ding\IEEEauthorrefmark{1}\IEEEauthorrefmark{2} and
Wen Xie\IEEEauthorrefmark{1}\IEEEauthorrefmark{4}}
\IEEEauthorblockA{\IEEEauthorrefmark{1}School of Computer Science and Engineering\\
Sichuan University Jinjiang College,
620860 Pengshan, China}
\IEEEauthorblockA{\IEEEauthorrefmark{2}School of Computer Science\\
Sichuan University,
610065 Chengdu, China}
\IEEEauthorblockA{\IEEEauthorrefmark{3}Email: greenhat1016@gmail.com}
\IEEEauthorblockA{\IEEEauthorrefmark{4}Corresponding Author Email: xwen@scu.edu.cn}}


\maketitle

\begin{abstract}
In this paper, a contrastive evaluation of massively parallel implementations of suffix tree and suffix array to accelerate genome sequence matching are proposed based on Intel Core i7 3770K quad-core and NVIDIA GeForce GTX680 GPU.
Besides suffix array only held approximately 20\%$\sim$30\% of the space relative to suffix tree, the coalesced binary search and tile optimization make suffix array clearly outperform suffix tree using GPU. Consequently, the experimental results show that multiple genome sequence matching based on suffix array is more than 99 times speedup than that of CPU serial implementation. There is no doubt that massively parallel matching algorithm based on suffix array is an efficient approach to high-performance bioinformatics applications.
\end{abstract}
\vspace{0.1in}

%
\IEEEpeerreviewmaketitle

\section{INTRODUCTION}
In recent years, modern multi-core and many-core architectures are revolutionizing high performance computing (HPC). As more and more processor cores are being incorporated into a single chip, the era of the many-core processor is coming. The emergence of many-core architectures, such as compute unified device architecture (CUDA)-enabled GPUs \cite{nvidia:motmanual} and other accelerator technologies (field-programmable gate arrays (FPGAs) and the Cell/BE), these technologies open up the the possibility of significantly reduce the runtime of many biological algorithms on commonly available and inexpensive hardware with more powerful high-performance computing power. Since the introduction of CUDA in 2007, more than 100 million computers with CUDA capable Graphics Processing Units have been shipped to end users. In the golden age of the GPU computing, with such a low barrier of entry, researchers all over the world have been engaged in developing new algorithms and applications to utilize the extreme floating point execution throughput of these GPUs.

Life science have emerged as a primary application area for the use of GPU computing. High-throughput techniques for DNA sequencing and gene expression analysis have led to an explosion of biological data. Prominent examples are the growth of DNA sequence information in NCBI's GenBank database and the growth of protein sequences in the UniProtKB/TrEMBL database. Furthermore, emerging next-generation sequencing technologies \cite{shendure2008next} have broken many experimental barriers to genome scale sequencing. Due to GPU performance grows faster than CPU performance, the use of GPUs in bioinformatics is a more appropriate strategy.

The suffix tree of a string is the compact trie of all its suffixes of the string, it's widely used in bioinformatics applications\cite{DBLP:books/cu/Gusfield1997}, e.g., MUMmer \cite{mummer3.0} and MUMmerGPU \cite{schatz2007high}. There are several approaches to construct the suffix tree in linear time \cite{788472}\cite{Ukkonen}\cite{McCreight:1976:SST:321941.321946}.  Nevertheless, with the growth of the reference sequence, the suffix tree will fall into the bottleneck of Dynamic Random Access Memory consumption. Because of the efficient usage of the cache memory and suffix array only take about 20\%$\sim$30\% of the space relative to suffix tree, the suffix array are sometimes preferred to the suffix tree in GPUs, i.e., genome sequence matching can be efficiently solved with suffix array. Meanwhile, in CPU there exist serial algorithms to construct suffix array in linear time \cite{karkkainen2006linear}\cite{Puglisi:2007:TSA:1242471.1242472}. In this paper, GPU implementations (suffix tree and suffix array) and optimization are presented to accelerate multiple genetic matching on two different platforms: multi-core (CPUs) and many-core (GPU). The GPU implementations show a tremendous performance boost, here the suffix array is more than 99 times speedup than that of CPU serial implementation and the suffix tree¡¯s speedup is approximately to 44-fold.

\section{Massively Parallel Processors}
\begin{figure}[H]
\centering
\includegraphics[width=0.50\textwidth]{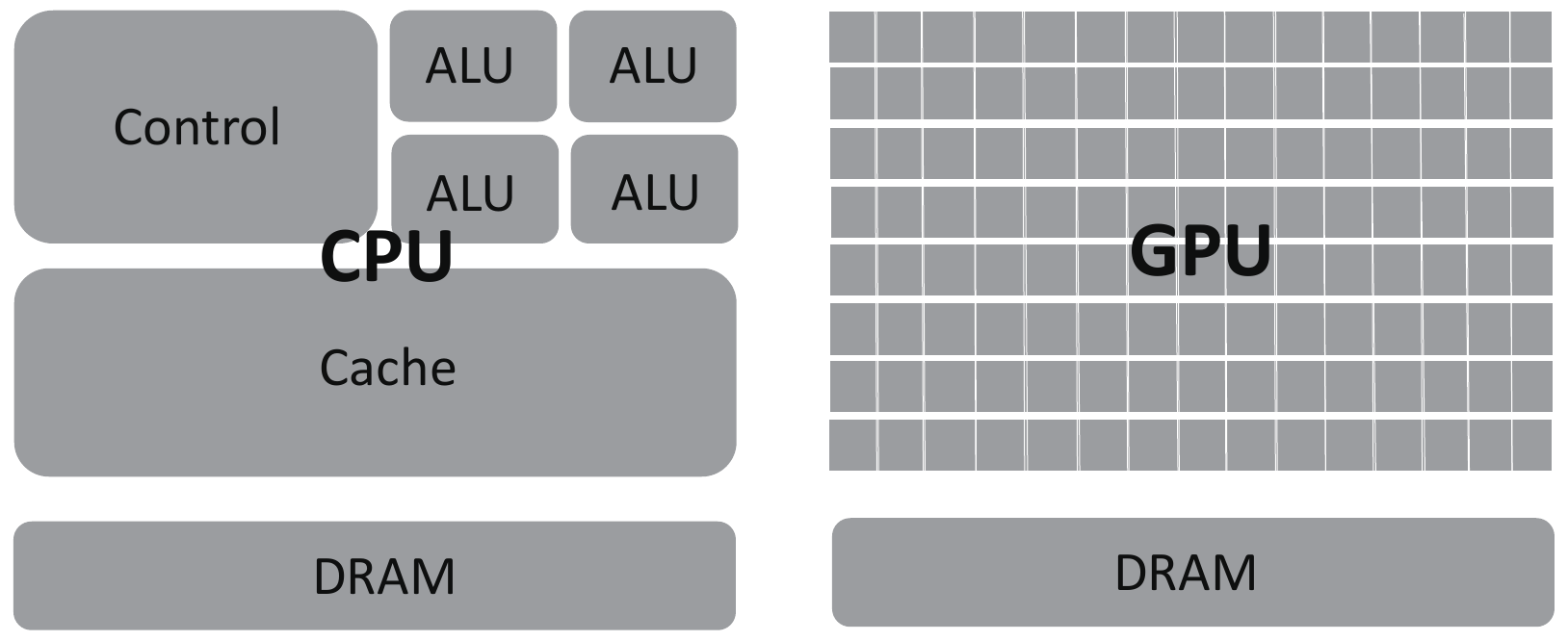}
\caption{CPUs and GPUs have fundamentally different design philosophies.}
\label{fig1}
\end{figure}

Since 2003, the semiconductor industry has settled on two main trajectories for designing microprocessor \cite{Hwu:2008:CC:1440404.1440423}. The many-core trajectory focuses more on the execution throughput of parallel applications, in contrast with the multicore trajectory (i.e. it maintains the execution speed of sequential programs while moving into multiple cores). The many-cores began as a large number of much smaller cores, and, once again, the number of cores doubles with each generation. A typical exemplar is NVIDIA$^\circledR$ GPU, each core is a in-order, heavily multi-threaded, single-instruction issue processor that shares its control and instruction cache with other cores. As illustrated in Fig. \ref{fig1}, the difference of the design philosophies in the two types of processors cause an enormous performance gap.

The original design philosophy of the GPUs is shaped by the fast growing video game industry, which exerts tremendous economic pressure for the ability to perform a massive number of floating-point calculations per video frame in advanced games. The hardware takes advantage of a large number of execution threads to find work to do when some of them are waiting for long-latency memory accesses, thus minimizing the control logic required for each execution thread. Small cache memories are provided to help control the bandwidth requirements of these applications so multiple threads that access the same memory data do not need to all go to the DRAM. As a result, much more chip area is dedicated to the floating-point calculations.

 Since a series of barriers and limitations, General-purpose programming using a graphics processing unit (GPGPU) was replaced by CUDA \cite{nvidia:motmanual}. CUDA programs no longer go through the graphics interface at all. Instead, a new general-purpose parallel programming interface on the silicon chip serves the requests of CUDA programs. In CUDA programming model, all threads in a grid execute the same kernel function, they rely on unique coordinates to distinguish themselves from each other and to identify the appropriate portion of the data to process. All threads are organized into a two-level hierarchy using unique coordinates -- $\textit{blockIdx}$ and $\textit{threadIdx}$. $\textit{gridDim}$ and $\textit{blockDim}$ provide the dimension of the grid and the dimension of each block respectively. According to the value of $\textit{gridDim}$ and $\textit{blockDim}$, dynamic partitioning of resources can result in subtle interactions between resource limitations, e.g., shared memory and registers capacity, the amount of blocks and threads, etc.

In modern software applications, program sections often exhibit a rich amount of data parallelism, a property allowing many arithmetic operations to be safely performed on program data structures in a simultaneous manner. The CUDA devices accelerate the execution of these applications by harvesting a large amount of data parallelism. Here, we just presented a short, informal discussion of GPU architecture and CUDA fundamentals \cite{schatz2007high}\cite{sanders2011cuda}\cite{kirk2010programming}. In recent years, a significant amount of new and interesting technologies have sprung up to efficiently solve problem in scientific research and commercial applications. There are several programming languages suitable for GPU programming, except for CUDA, the most common being OpenCL \cite{gaster2012heterogeneous}, OpenACC, and C++ AMP \cite{gregory2012c++}.

\section{Suffix Array Construction}

\begin{figure}
\centering
\includegraphics[width=0.40\textwidth]{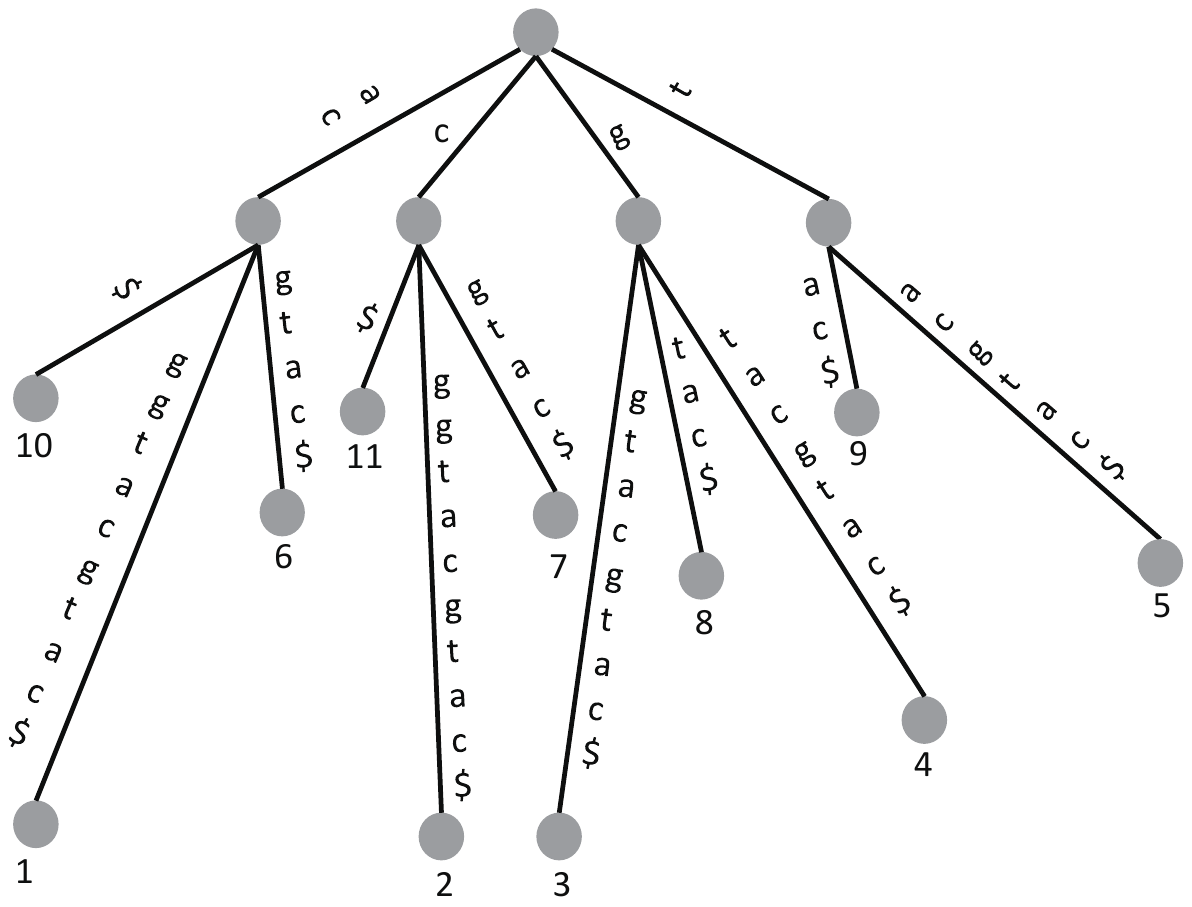}
\caption{The suffix tree of reference sequence S = $acggtacgtac$}
\label{fig2}
\end{figure}

\begin{figure}[H]
\centering
\includegraphics[width=0.35\textwidth]{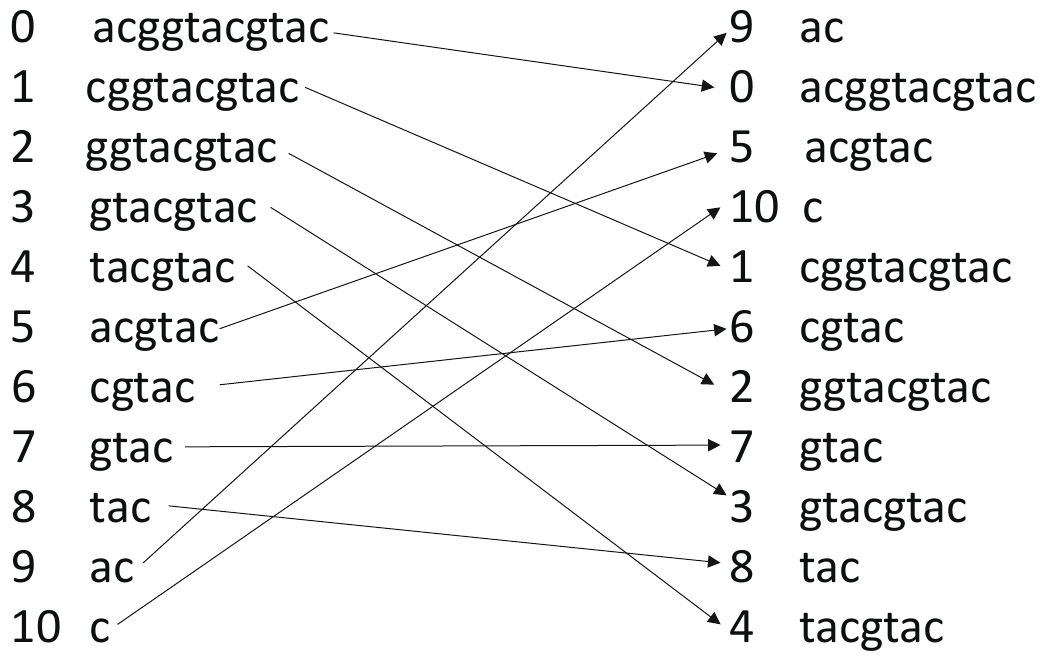}
\caption{For a reference sequence S = $acggtacgtac$, all the suffixes are added into the array and sorted lexicographically.}
\label{fig3}
\end{figure}

For convenience, $n$ is indicated as the length of reference sequence and $m$ is denoted as the length of query sequence. Given a reference sequence $S = s_{1}s_{2}...s_{n}$. For $i = 1, 2, ..., n$, every $S(i, n)$ is a suffix of $S$.  Initially, for the four nucleic acid bases include adenine, guanine, thymine, and cytosine that makes up DNA, define $\Sigma = \{a,c,g,t\}$. We shall label the suffixes according to the location of the starting character, that is, $S_{i} = S(i, n)$. For example, if $S$ = $acggtacgtac$, $S_{2}$ = $cggtacgtac$ and $S_{8}$ = $gtac$, just like the left-hand side of Fig. \ref{fig3}.
A suffix tree of $S$ of length $n$ is a tree with the following properties: 1) Each tree edge is labeled by a subsequence of $S$. 2) Each internal node has at least two children. 3) For $1\leq i \leq n$, each $S_{i}$ has its corresponding labeled path from root to a leaf. 4) No edges branching out from the same internal node can start with the same character. The typical suffix tree is illustrated as Fig. \ref{fig2}.

The suffix tree can be constructed from the reference sequence in $O(n)$ linear time \cite{Ukkonen}. Nevertheless, all the significant features provided by suffix tree are offered at the cost of an important drawback, related to the amount of space that is required to store this index structure, which can be nearly 20-fold with respect to the initial reference size. Afterwards, the suffix tree can be transmitted into flatten tree consisting of an array of edges and the result of the GPU accelerate genome sequence alignment based on flatten tree is efficient.

Still, when compared to suffix trees, the suffix arrays are regarded as a more space-efficient implementation. This structure can be served as an array of integers representing the start position of every lexicographically ordered suffix of a string. After $S$ is lexicographically sorted, as illustrated in Fig. \ref{fig3}, lexicographical suffix indexes are filled into the suffix array $SA$. \ref{fig3}. As Table \ref{table1} indicates, the suffix array is defined as $SA$.

\begin{table}[H]

\centering
\caption{The sorted suffix index table for S = $acggtacgtac$}
\label{table1}
\begin{tabular}{|c|c|c|c|c|c|c|c|c|c|c|c|} \hline
Index&0&1&2&3&4&5&6&7&8&9&10\\ \hline
SA&9&0&5&10&1&6&2&7&3&8&4\\ \hline

\end{tabular}
\end{table}

The most straightforward way to construct the suffix array is to simply create an array with all the suffix elements placed in ascending order and then apply a sorting algorithm to properly sort the suffixes. We can utilize Difference Cover Modulo 3 (DC3) algorithm to implementation in $O(n)$ time.

A linear time suffix array construction algorithm, namely DC3. It takes the following $2/3$-recursive divide-and-conquer approach:

\noindent $\bold{1}.$  Construct the suffix array of the suffixes starting at position $i\ mod\ 3 \neq 0$.

For $k = 0,1,2$, define $B_{k} = \{i \in [0,n] \ | \ i \ mod \ 3 = k\}$. Let $C = B_{1} \bigcup B_{2}$ be the set of sample positions and $S_{C}$ the set of sample suffixes. For a reference sequence $S$ = $acggtacgtac$, $B_{1}$ = $\{1,4,7,10\}$, $B_{2}$ = $\{2,5,8\}$ and $C$ = $\{1,4,7,10,2,5,8\}$. For $k = 1,2$, construct the strings $R_{1} =$ [$cgg$] [$tac$] [$gta$] [$c00$], $R_{2} =$ [$ggt$] [$acg$] [$tac$] and $R = R_{1} \bigodot R_{2} =$ [$cgg$] [$tac$] [$gta$] [$c00$] [$ggt$] [$acg$] [$tac$]. Radix sort\cite{Cormen:2009:IAT:1614191} the characters are of $R$ and rename them with ranks to obtain $rank(S_{i})$, which is illustrated by Table \ref{table2}.

\begin{table}[H]
\centering
\caption{For $S = acggtacgtac$, the ranks of the sorted sample suffixes}
\label{table2}
\begin{tabular}{|c|c|c|c|c|c|c|c|c|c|c|c|}\hline

 $i$ & 0 & 1 & 2 & 3 & 4 & 5 & 6 & 7 & 8 & 9 & 10 \\ \hline
  $rank(S_{i})$ & $\bot$ & $3$ & $4$ & $\bot$ & $7$ & $1$ & $\bot$ & $5$ & $6$ & $\bot$ & $2$\\ \hline

\end{tabular}
\end{table}

\noindent $\bold{2}.$  Construct the suffix array of the remaining suffixes using the result of the first step.

Represent each non-sample suffix $S_{i} \in S_{B_{0}}$ with the pair ($S_{i}, rank(S_{i+1})$). For all $i,j \in B_{0}$, $S_{i} \leq S_{j} \Leftrightarrow (S_{i}, rank(S_{i+1})) \leq (S_{j}, rank(S_{j+1}))$. The pairs are then radix sorted. For $S = acggtacgtac$, $B_{0} = \{0,3,6,9\}$. Because of $(a,2) \leq (a,3) \leq (c,5) \leq (g,7)$, $S_{9} \leq S_{0} \leq S_{6} \leq S_{3}$.

\noindent $\bold{3}.$  Merge the two suffix arrays into one using a standard comparison-based merging.To compare suffix $S_{i}\in S_{C}$ with $S_{j}\in S_{B_{0}}$, There are two cases£º

1) $i\in$ $B_{1}$, $S_{i}\leq S_{j}\Leftrightarrow$ $(t_{i}$,$rank(S_{i+1}))$ $\leq$ $(t_{j}$,$rank(S_{j+1}))$

2) $i\in$ $B_{2}$, $S_{i}\leq S_{j}\Leftrightarrow$ $(t_{i}$,$t_{i+1}$,$rank(S_{i+2}))$ $\leq$ $(t_{j}$,$t_{j+1}$,
$rank(S_{j+2}))$

\begin{table}[H]
\centering
\caption{For $S = acggtacgtac$, the ranks of all suffixes}
\label{table3}
\begin{tabular}{|c|c|c|c|c|c|c|c|c|c|c|c|}\hline

 $i$ & 0 & 1 & 2 & 3 & 4 & 5 & 6 & 7 & 8 & 9 & 10 \\ \hline
  $rank(S_{i})$ & $2$ & $5$ & $7$ & $9$ & $11$ & $3$ & $6$ & $8$ & $10$ & $1$ & $4$\\ \hline

\end{tabular}
\end{table}

Consequently, the time complexity of algorithm DC3 is $O(n)$. Several parallel and hierarchical memory models of computations are presented in \cite{karkkainen2006linear}\cite{Sintorn:2008:FPG:1412749.1412828}.

\section{Indexed Parallel Searching}

After the suffixes are sorted, the indexed structure is used for searching query sequence $P$. For a reference sequence $S$ = $acggtacgtac$, as described in table \ref{table1} and Fig. \ref{fig3}, the first suffix $SA_{0}$ is denoted as $ac$, $SA_{4}$ = $cggtacgtac$ and $SA_{6}$ = $ggtacgtac$. If $P$ = $c$, we would find them in locations 3 to 5 in the sorted suffix index table. Similarly, $P$ = $a$ appears at three locations, namely 9, 0, 5 in reference sequence $S$. This just goes to illustrate that for every query sequence $P$ which appears in $S$, there is a range which it appears in the sorted suffix index table. The range is defined by its left boundary, denoted as $LB$, and its right boundary, denoted as $RB$. For $P$ = $a$, $LB$ = 0 and $RB$ = 2. For $P$ = $c$, $LB$ = 3 and $RB$ = 5. For $P$ = $ggtac$, $LB$ = $RB$ = 6. Therefore, our job is to search for $LB$ and $RB$.

Given a query sequence $P$, in general, binary search algorithm is an ideal strategy. During the binary search, we compare $P$ with a suffix $S_{i}$ in the suffix array. There are two possibilities: 1) $P$ is a prefix of $S_{i}$ and 2) $P$ is not a prefix of $S_{i}$.

\noindent $\bold{1}.$ If $P$ is a prefix of $S_{i}$, $LB$ may be $i$ or to the left of $i$ and $RB$ may only be $i$ or the right of $i$. Thus search both left and right of $S_{i}$ is necessary.

\noindent $\bold{2}$ If $P$ is not a prefix of $S_{i}$ and $P < S_{i}$, both $LB$ and $RB$ must be to the left of $S_{i}$ if $P$ occurs in $S$.

\noindent $\bold{3}$ If $P$ is not a prefix of $S_{i}$ and $P > S_{i}$, both $LB$ and $RB$ must be to the right of $S_{i}$ if $P$ occurs in $S$.

If $P$ is a prefix of $S_{i}$ and $P$ is not a prefix of $S_{i-1}$, then $LB = i$. Similarly, $RB = i$ if $P$ is a prefix of $S_{i}$ and $P$ is not a prefix of $S_{i+1}$. If $S$ = $acggtacgtac$, for $P$ = $c$, $LB = 3$ and $RB = 5$. Assume that $P$ = $tac$, $LB$ = 9 and $RB$ = 10. Both $LB$  and $RB$ can be assigned using binary search algorithm.

\begin{algorithm}
\small
\caption{Kernel function \textit{cudaGeneBinSearch}: GPU threads parallel matching multiple genome sequence using binary search algorithm} \label{alg1}
\begin{algorithmic}

   \item[1] //$geneSubsequence[thd]\ \ is\ \ one\ \ of\ \ genome\ \ subsequences$
   \item[2] thd = \textit{threadIdx.x}$+$\textit{blockDim.x}$*$\textit{blockIdx.x};	
    \item[3]//$assign$ $the$ $boundaries$ $to$ $registers$
    \item[4] L = left, R = right
    \item[5] \_\_shared\_\_ midSuffix[segLength], querySequence[segLength]
    \item[6]//$an$ $O(m\log{n})$ $search$ $algorithm$ $to$ $determine$ $LB$
    \item[7] \ \ $\bold{while}($R $>$ L $+$ $1)$
    \item[8] $\{$
	      \item[9]\ \ \ \ pivot = (L + R) $\gg$ 1
	    \item[10] \ \ \ \ $\bold{do}$\ $\{$
	     \item[11] \ \ \ \ adjust\ \ middleSuffxLength\ \ and\ \ querySequenceLength
		 \item[12]  \ \ \ \ //$update$\ \ $and$\ \ $extract$\ \ $trailing$\ \ $pair$\ \ $bases$
          \item[13]  \ \ \ \ //$send$\ \ $extracted$\ \ $segment$\ \ $to$\ \ $shared$\ \ $memory$
		 \item[14] \ \ \ \ midSuffix $=$ extract\ segment\ from\ S[SA[pivot]]
	     \item[15] \ \ \ \ querySequence $=$ extract\ segment\ from\ QuerySeq[thd]

		 \item[16]\ \ \ \ $\bold{if}$(midSuffix $!=$ querySequence)\ \ $\bold{break}$			
		 \item[17]\ \ \ $\}\bold{while}($middleSuffxLength $> 0$ \&\& querySequenceLength $> 0)$

           \item[18]\ \ \ \ $\bold{if}$(midSuffix $\leq$ querySequence)
         \item[19] \ \ \ \ \ \ \ \ $\bold{then}$\ \ R = pivot
         \item[20]\ \ \ \ $\bold{else}$
                \item[21] \ \ \ \ \ \ \ \ $\bold{then}$\ \ L = pivot
	 \item[22] \}
        \item[23] $LB$ = R
        \item[24] L = left, R = right
        \item[25]//$an$ $O(m\log{n})$ $search$ $algorithm$ $to$ $determine$ $RB$ $which$ $similar$ $to$ $LB$

        \item[26] \ \ $\bold{while}($R $>$ L $+$ $1)$
    \item[27] $\{$
	      \item[28]\ \ \ \ pivot = (L + R) $\gg$ 1
	    \item[29] \ \ \ \ $\bold{do}$\ $\{$
	     \item[30] \ \ \ \ adjust\ \ middleSuffxLength\ \ and\ \ querySequenceLength
		 \item[31]  \ \ \ \ //$update$\ \ $and$\ \ $extract$\ \ $trailing$\ \ $pair$\ \ $bases$
          \item[32]  \ \ \ \ //$send$\ \ $extracted$\ \ $segment$\ \ $to$\ \ $shared$\ \ $memory$
		 \item[33] \ \ \ \ midSuffix $=$ extract\ segment\ from\ S[SA[pivot]]
	     \item[34] \ \ \ \ querySequence $=$ extract\ segment\ from\ QuerySeq[thd]

		 \item[35]\ \ \ \ $\bold{if}$(midSuffix $!=$ querySequence)\ \ $\bold{break}$			
		 \item[36]\ \ \ $\}\bold{while}($middleSuffxLength $> 0$ \&\& querySequenceLength $> 0)$

           \item[37]\ \ \ \ $\bold{if}$(midSuffix $<$ querySequence)
         \item[38] \ \ \ \ \ \ \ \ $\bold{then}$\ \ R = pivot
         \item[39]\ \ \ \ $\bold{else}$
                \item[40] \ \ \ \ \ \ \ \ $\bold{then}$\ \ L = pivot
	 \item[41] \}

        \item[42] $RB$ = L
        \item[43] $\_\_$syncthreads()
        \item[44] res[thd $\ll$ 1] = LB
        \item[45] res[thd $\ll$ 1 + 1] = RB
\end{algorithmic}
\end{algorithm}

The algorithm can be modified to implement multiple query sequences matching concurrently, the pseudocode is depicted in Alg.\ref{alg1}. In the kernel function \textit{$cudaGeneBinSearch$}, $thd$ is a private register with respect to a thread respectively, which denote as thread index. All query sequences are organized to a string array QuerySeq so that the exclusive thread can be mapped into QuerySeq[$thd$]. The reference sequence is constructed as suffix array so as to the indexed structure enable all threads to string matching simultaneously by means of binary search algorithm supported by GPU. The boundary of the reference sequence can be determined by variables, that is, $left$ and $right$.

Assume that the length of the reference sequence is $n$ and the length of the query sequence is $m$, the algorithm take $O(m\log{n})$ time to determine $LB$ or $RB$. ($LB$,$RB$) is the matching results with respect to QuerySeq[$thd$]. If ($LB$,$RB$) is $NULL$, there is no suffix similar to query sequence. From alg.\ref{alg1}, when $pivot$ is determined by $(L + R) \gg 1$, we extract a part of sequence segments from global memory to shared memory until middleSuffixLength $\leq 0$ or querySequenceLength $\leq 0$, i.e., S[SA[$pivot$]]$\rightarrow$ midSuffix and QuerySeq[$thd$] $\rightarrow$ querySequence. Global memory is large but slow, whereas the shared memory is small
but fast. In 2011, G. Encarnaijao, N. Sebastiao, and N. Roma \cite{5999806} proposed an analogical approach but they didn't exactly discuss the range of results. Their method just get one matching sequence, however, our job is to search for $LB$ and $RB$, this is, to get all satisfied results. But above all, we partition the data into subsets called tiles such that each tile fits into the shared memory. It is an effective strategy for achieving high performance in virtually all types of parallel computing systems. In 43rd pseudocode line, $\_\_$syncthreads() function be viewed as a command to be waiting all threads execution in a block, namely synchronization.
Only when all threads execute over here, the result of the boundaries ($LB$,$RB$) with respect to QuerySeq[$thd$] can be assigned and transfered to host.

\begin{table*}
\caption{The benchmark test of suffix tree and suffix array indexed search algorithms in multi-core CPUs}
\begin{center}
SA: Suffix Array  \ \ ST: Suffix Tree
\end{center}
\centering

\begin{tabular}{|c|c|c|c|c|c|c|c|c|} \hline
Genome Sequences	    &Serial SA	&Serial ST	&2 threads SA	&2 threads ST	&4 threads SA	&4 threads ST &8 threads SA	&8 threads ST\\ \hline
512	    &0.002	&0.002	&0.001	&0.001	&0.001	&0.001	&0.001	&0.001\\ \hline
1024	&0.005	&0.004	&0.003	&0.002	&0.002	&0.001	&0.001	&0.001\\ \hline
2048	&0.01	&0.007	&0.006	&0.004	&0.003	&0.002	&0.002	&0.001\\ \hline
4096	&0.04	&0.03	&0.022	&0.017	&0.013	&0.009	&0.007	&0.005\\ \hline
8192	&0.075	&0.056	&0.042	&0.032	&0.024	&0.018	&0.013	&0.01\\ \hline
16384	&0.25	&0.188	&0.141	&0.105	&0.079	&0.059	&0.044	&0.033\\ \hline
32768	&0.72	&0.54	&0.405   &0.304	&0.228	&0.171	&0.128	&0.096\\ \hline
65536	&1.22	&0.915	&0.686	&0.515	&0.386	&0.29	&0.217	&0.163\\ \hline
131072	&2.35	&1.762	&1.322	&0.991	&0.744	&0.558	&0.418	&0.314\\ \hline
262144	&5.01	&3.758	&2.818	&2.114	&1.585	&1.189	&0.892	&0.669\\ \hline
524288	&11.67	&8.753	&6.564	&4.923	&3.692	&2.769	&2.077	&1.558\\ \hline
1048576	&30.6	&22.95	&17.213	&12.909	&9.682	&7.262	&5.446	&4.085\\ \hline
2097152	&55.11	&41.33	&30.999	&23.25	&17.437	&13.078	&9.808	&7.356\\ \hline

\end{tabular}

\end{table*}

\begin{table*}
\caption{a contrastive evaluation of MUMmerGPU, indexed search based on Suffix Tree and Suffix Array in NVIDIA GeForce GTX680 GPU}
\begin{center}MI: MUMmerGPU Input Time  \ \  MK: MUMmerGPU Kernel Time \ \ MO: MUMmerGPU Output Time \ \ MT: MUMmerGPU Total Time
 \\    STI: Suffix Tree Input Time \ \ STK: Suffix Tree Kernel Time \ \ STO: Suffix Tree Output Time \ \ STT: Suffix Tree Total Time
 \\    SAI: Suffix Array Input Time \ \ SAK: Suffix Array Kernel Time \ \ SAO: Suffix Array Output Time \ \ SAT: Suffix Array Total Time\end{center}

\centering
\begin{tabular}{|c|c|c|c|c|c|c|c|c|c|c|c|c|} \hline
Genome Sequences	&MI	 &MK	&MO	&MT &STI &STK	&STO	&STT	&SAI   &SAK	&SAO	&SAT\\ \hline

512	    &0.1	    &0.001	&0.00005	&0.1011  &0.088	&0.0002	&0.000025	&0.0882   &0.0082	&0.0003	&0.000025	&0.0085\\ \hline
1024	&0.1	    &0.001	&0.00006	&0.1011  &0.089	&0.0004	&0.000026	&0.0894   &0.0082	&0.0004	&0.000026	&0.0086\\ \hline
2048	&0.12	&0.002	&0.00007	&0.1221  &0.089	&0.0005	&0.00003	   & 0.0895  & 0.0083	&0.0005	&0.00003	    &0.0088\\ \hline
4096	&0.15	&0.0026	&0.00008	&0.1527  &0.089	&0.0007	&0.000043	&0.0897   &0.0083	&0.0006	&0.000045	&0.0089\\ \hline
8192	&0.18	&0.0043	&0.00011	&0.1844  &0.09	&0.0009	&0.00064	    &0.091    &0.0084	&0.0006	&0.00006	    &0.0091\\ \hline
16384	&0.21	&0.0069	&0.00035	&0.2173  &0.09	&0.0021	&0.00089	    &0.0922   &0.0085	&0.0013	&0.000087	&0.0099\\ \hline
32768	&0.26	&0.0096	&0.00062	&0.2702  &0.09	&0.0063	&0.00023	    &0.0965   &0.0087	&0.0063	&0.00024	   & 0.0152\\ \hline
65536	&0.34	&0.032	&0.00081	&0.3728  &0.09	&0.0081	&0.0045	    &0.0985   &0.009	&0.0116	&0.00049	    &0.0211\\ \hline
131072	&0.41	&0.043	&0.002	&0.455   &0.09	&0.018	&0.0072	    &0.1087   &0.01	&0.0398	&0.00074	   & 0.0505\\ \hline
262144	&0.48	&0.063	&0.0065	&0.5495  &0.91	&0.044	&0.0099	    &0.1449   &0.02	&0.0579	&0.00099	   & 0.0789\\ \hline
524288	&0.63	&0.081	&0.0083	&0.7193  &0.133	&0.1781	&0.024	    &0.3135   &0.031	&0.095	&0.0024	   & 0.13284\\ \hline
1048576	&0.8	    &0.21	&0.02	&1.03   & 0.253	&0.22	&0.05	    &0.523    &0.045	&0.2716	&0.0048	   & 0.3214\\ \hline
2097152	&1.05	&0.45	&0.14	&1.64    &0.511	&0.33	&0.088	    &0.929    &0.0695	&0.4762	&0.0088	    &0.5545\\ \hline

\end{tabular}

\end{table*}

\begin{figure*}
\centering
\includegraphics[width=1.0\textwidth]{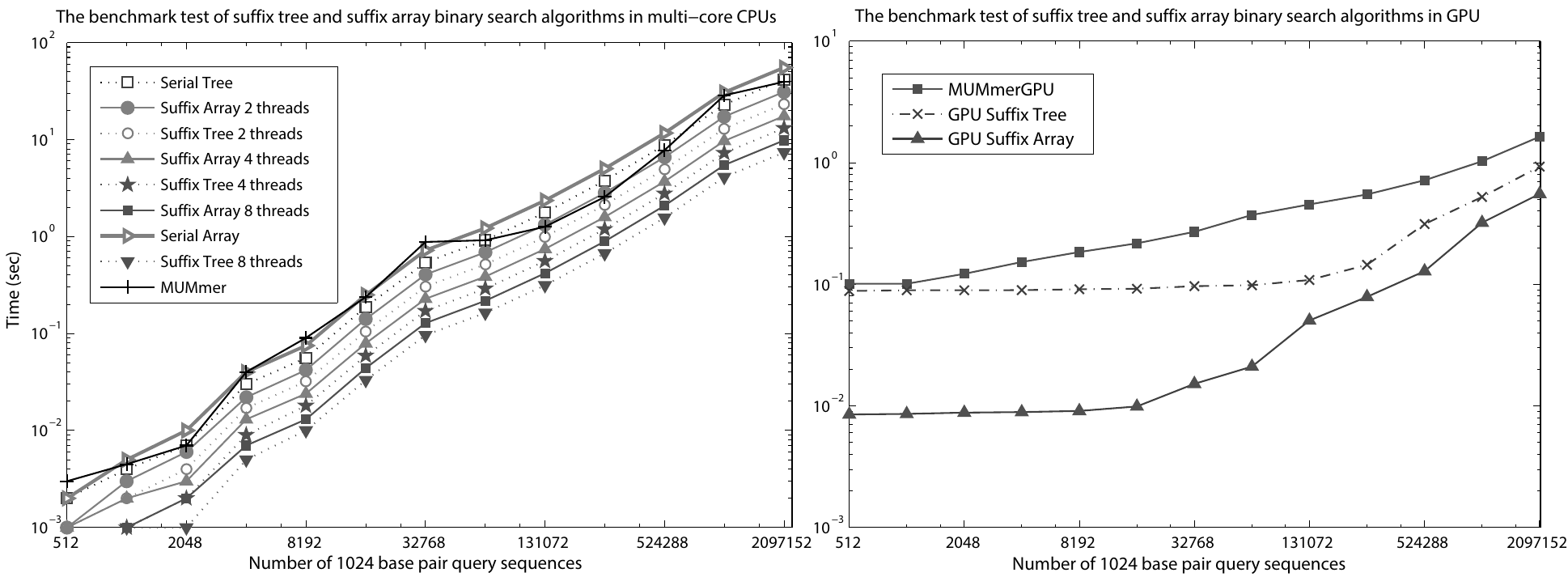}
\caption{Benchmark test of suffix tree and suffix array indexed based search algorithms in multi-core CPUs and GPU.}
\label{fig6}
\end{figure*}
\section{Results}

The indexed based search algorithms were evaluated in a computer composed of an Intel Core i7 3770K quad-core processor, running at 3.5GHz, with 1600MHz and 16GB DRAM. It's worth mentioning that the graphics card is NVIDIA GeForce GTX 680 that is Kepler GPU architecture, with 1536 stream processors running at 6008MHz and 2GB RAM.

We extracted DNA sequence from NCBI Nucleotide to evaluate the performance of the previously described algorithms. The reference sequence, which was used to build the indexed structure (suffix tree and suffix array), corresponds to the first $10^7$ nucleotides of the NT\_167186.1 Homo sapiens chromosome 1 genomic contig. Query sequences we used are 1024 nucleotides long, which derive from a mix of the DNA sequences extracted from the NT\_167186.1 Homo sapiens chromosome 1 genomic contig and the NT\_039173.8 Mus musculus strain C57BL6J chromosome 1 genomic contig. Several sets of query sequences were used in the experimental test, each one consists of a different number of query sequences, ranging from 512 to 2097152 query sequences. No matter what the number of query sequences in a specific set, their containing query sequences all are 1024 nucleotides long.

In the preliminary estimation, in order to evaluate the best performance, we evaluated the parallel implementations of genome sequence search algorithms based either on the suffix tree and on the suffix array index in two different parallel platforms. In a homogeneous multi-core CPU, the algorithms are executed using OpenMP \cite{pacheco2011introduction} to support the parallel execution of 2, 4 and 8 concurrent threads. Simultaneously it also provided a comparative evaluation with highly efficient CPU based software named MUMmer\cite{mummer3.0}.

\begin{figure}
\centering
\includegraphics[width=0.5\textwidth]{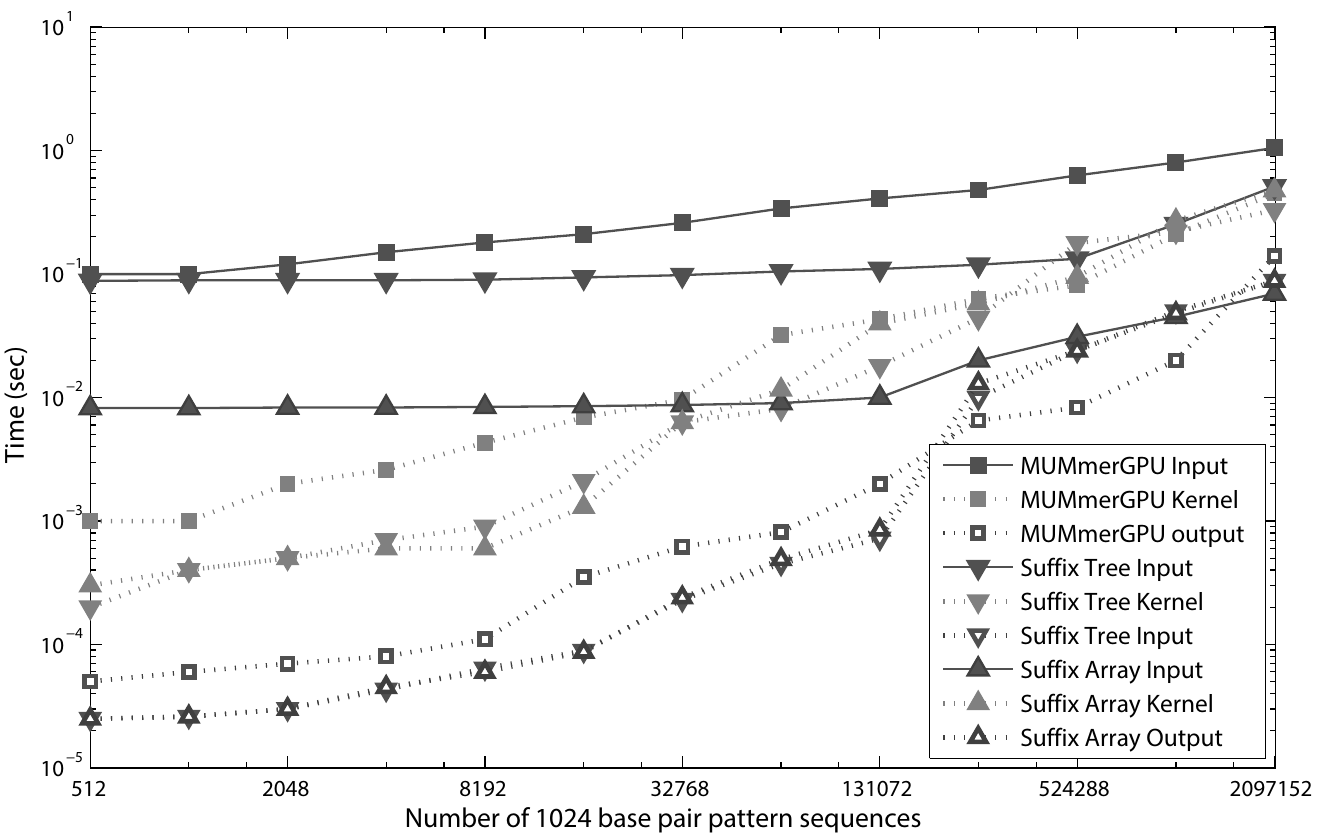}
\caption{kernel execution and Communication time for the algorithms in the NIVIDIA GeForce GTX 680 GPU.}
\label{fig7}
\end{figure}

The results is illustrated in the left side of Fig. \ref{fig6}.
Although suffix tree take $O(m)$ time to search and suffix array take $O(mlog{n})$ time to search, due to the more coalescence exist in the indexed search algorithm based on suffix array, when accessing memory, the asymptotic runtime correspond to the suffix array is slightly greater than that of the suffix tree. Because of pointer skip or index skip to next waiting-for-matched character in indexed search algorithm based on suffix tree, suffix tree is not coalescing clearly. It also observed that the parallel implemented suffix tree and suffix array are significantly faster than MUMmer.

\begin{figure}
\centering
\includegraphics[width=0.50\textwidth]{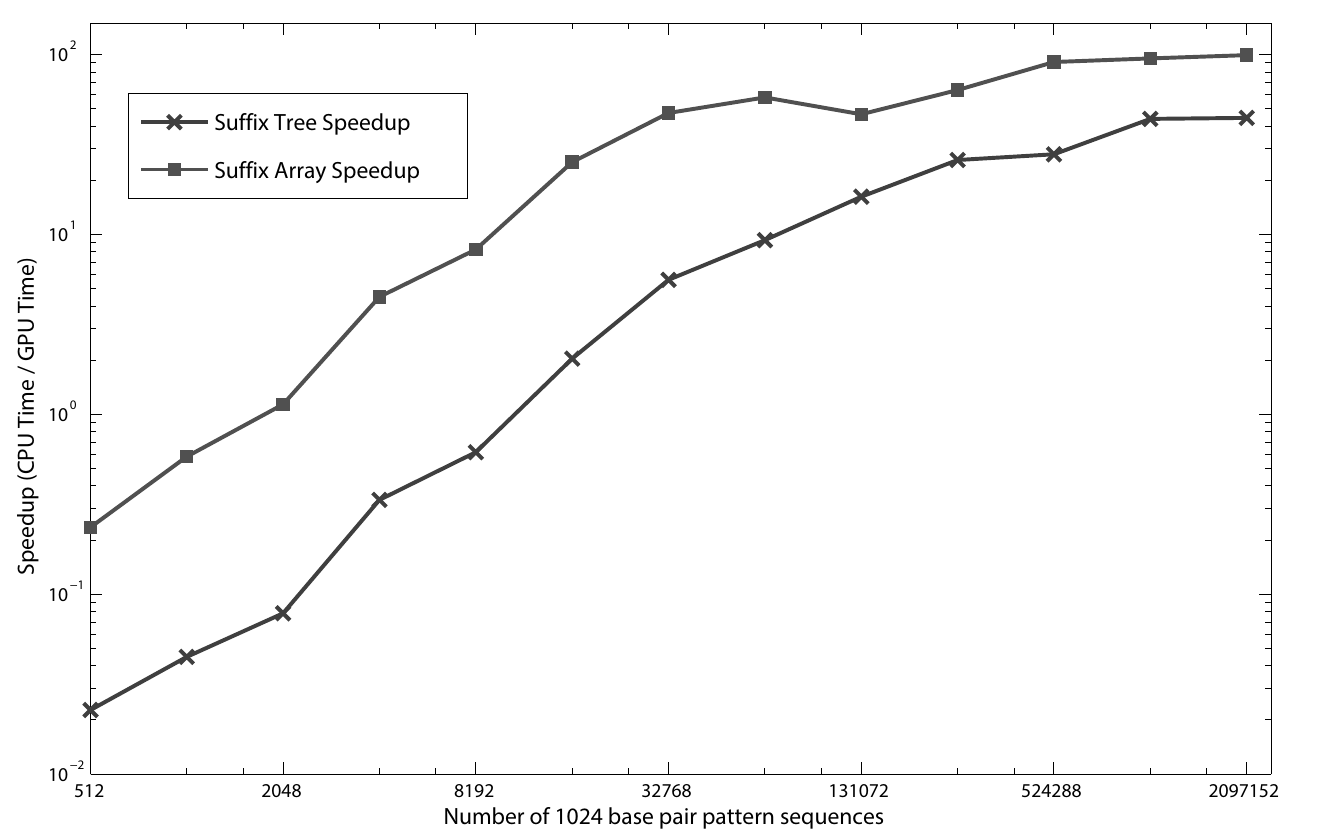}
\caption{The speedup ratio of the algorithms between CPU and GPU.}
\label{fig8}
\end{figure}

Furthermore, the performance of the parallel algorithms was evaluated by NVIDIA GeForce GTX 680. The obtained results are depicted in the right side of Fig. \ref{fig6}. These results correspond to the total runtime of the search algorithms include communication time (transfer data between host and device) and kernel execution time. The performance of the separated components are portrayed in Fig. \ref{fig7}. Just like MUMmer based on CPU, there is a genome sequence alignment tool based on GPU architecture, namely MUMmerGPU\cite{schatz2007high}. It's worth noting that this figure not contain a comparison with CUDASW+, because of maximum reference size of about $64*10^3$ base pairs is the hugely limit factor. Since the indexed structure must be transferred from the CPU host memory to the GPU device memory, the input time is critical to all index-based search algorithms. Indeed, when the number of query sequences to be searched is very small, the data input time occupy the vast majority of total time. Nevertheless, for a larger number of query sequences, the throughput of the highly parallel implementations by GPU amortized the input time. The GPU implementations show a tremendous performance boost, when the number of query sequences is about 2097152, the suffix array is more than 99 times speedup than that of CPU serial implementation and the suffix tree's speedup is approximately to 44-fold. The more details are observed in Fig. \ref{fig8}.

Unlike what happened in CPU implementations, due to the achieved performance in accordance with memory accesses, tile optimization partition local data into shared memory, which reduce the access of global memory. On the other hand, because of the suffix array only held about 20\% $\sim$ 30\% space consumption than suffix tree, when the whole indexed structure is transfered to GPU, the suffix array apparently take fewer input time than suffix tree. The experimental results show that GPU implementations clearly favor suffix array. The reason for this situation is not only the more coalescence of this algorithm and its more efficient use of the shared memory, but is also the space consumption of the suffix array is much smaller than that of the suffix tree, which makes the suffix array implementation to always present a much lower transfer time from the host to the GPU device. From the obtained results it can be observed that the runtime of MUMmerGPU were consistently higher than the implemented suffix tree and suffix array, that is, the indexed search algorithm with respect to suffix array using GPU, even suffix tree, is efficient method to high performance bioinformatics applications.

\section{Conclusion}
This paper proposed a comparative evaluation of suffix tree and suffix array, which are extra applicable for accelerating DNA sequence matching. These indexed structures were thoroughly compared using two different parallel platforms: multi-core (i7 3770K quad-core) and many-core (NVIDIA GeForce GTX680 GPU).

These observations reveal that suffix array is slightly greater than suffix tree though the asymptotic search time of the suffix array ($O(mlogn)$) is far higher than that of the suffix tree ($O(m)$) in multi-core platform, due to the more coalescence exist in the indexed search algorithm based on suffix array. Moreover, because of the tile optimization and fewer transfer time in suffix array relative to suffix tree, the obtained results show that suffix array clearly outperform suffix tree in many-core platform. According to the results, there are convincing reasons to believe that massively parallel matching based on suffix array using GPU is an efficient approach to high-performance bioinformatics applications.

\section*{Acknowledgment}
This work was completely supported and funded by Sichuan University Jinjiang College. All authors read and approved the final manuscript.



%
\bibliographystyle{IEEEtran}
\bibliography{example}

\end{document}